\documentclass{article}
\usepackage{spconf,amsmath,graphicx,hyperref}
\usepackage{amsmath}
\usepackage{amsbsy}
\usepackage{amssymb}
\usepackage{amsfonts}
\usepackage{amsthm}
\usepackage{color}

\providecommand{\vh}{\mathbf{h}}

\providecommand{\vx}{\mathbf{x}}
\providecommand{\vy}{\mathbf{y}}

\usepackage{cite}
\usepackage{amssymb}
\usepackage{booktabs}
\usepackage{xcolor}
\usepackage{multirow}
\usepackage{subcaption}
\usepackage[capitalise]{cleveref}
\usepackage{lineno}

\usepackage[toc,page]{appendix}

\title{Diffusion Algorithm for Metalens Optical Aberration Correction}
\name{
\begin{tabular}{c}
Harshana Weligampola$^1$, Yuanrui Chen$^1$, 
Weiheng Tang$^1$, \\
Qi Guo$^1$, Stanley H. Chan$^1$
\end{tabular}
\thanks{The work was supported by Samsung Research America Global Outreach and U.S. National Science Foundation award CCF\_2431505.}
}

\address{
$^1$Elmore Family School of Electrical and Computer Engineering, Purdue University\\
}
\begin{document}
\maketitle
\begin{abstract}
Metalenses offer a path toward creating ultra-thin optical systems, but they inherently suffer from severe, spatially varying optical aberrations, especially chromatic aberration, which makes image reconstruction a significant challenge. This paper presents a novel algorithmic solution to this problem, designed to reconstruct a sharp, full-color image from two inputs: a sharp, bandpass-filtered grayscale ``structure image'' and a heavily distorted ``color cue'' image, both captured by the metalens system. Our method utilizes a dual-branch diffusion model, built upon a pre-trained Stable Diffusion XL framework, to fuse information from the two inputs. We demonstrate through quantitative and qualitative comparisons that our approach significantly outperforms existing deblurring and pansharpening methods, effectively restoring high-frequency details while accurately colorizing the image.
\end{abstract}
\begin{keywords}
spatially varying deblurring, metalens, optics, aberrations, diffusion
\end{keywords}
\section{Introduction}
\label{sec:intro}

The proliferation of metalenses has created an unprecedented opportunity to develop ultrathin optical elements that, in specific settings, can approach the performance of traditional refractive lens systems \cite{chakravarthula_tog_2023_thinwidelens, metasurface_review, WFoV, Metaoptical_zoom}. At the core of metalenses lies an array of nanoscale structures with engineered phase profiles \cite{Brookshire:24, jumping} that modify the phases of incident waves as they exit the metasurface. Through careful design, researchers have envisioned a new generation of optical systems with extremely compact form factors. In both academia and industry \cite{metastatinsight}, metalenses are attracting significant attention.

While metalenses possess many desirable features, one of the most challenging aspects lies surprisingly not in the hardware, but in postprocessing image reconstruction. Due to the limited bandwidth that a metalens can support, it inherently suffers from optical aberrations, with chromatic aberration being particularly problematic. In practice, this means that if a metalens performs well at one wavelength, it generally performs poorly at adjacent wavelengths \cite{rgbeyepiece}. Although significant efforts are underway to develop wideband metalenses \cite{chen2018broadband}, the current burden largely falls on image reconstruction algorithms. However, reconstruction is highly challenging, as the aberrations are spatially varying and produce a wide spread of blur across the field of view, making the underlying deconvolution extremely difficult, if not infeasible.

\begin{figure}[t]
\centering
\includegraphics[width=\linewidth]{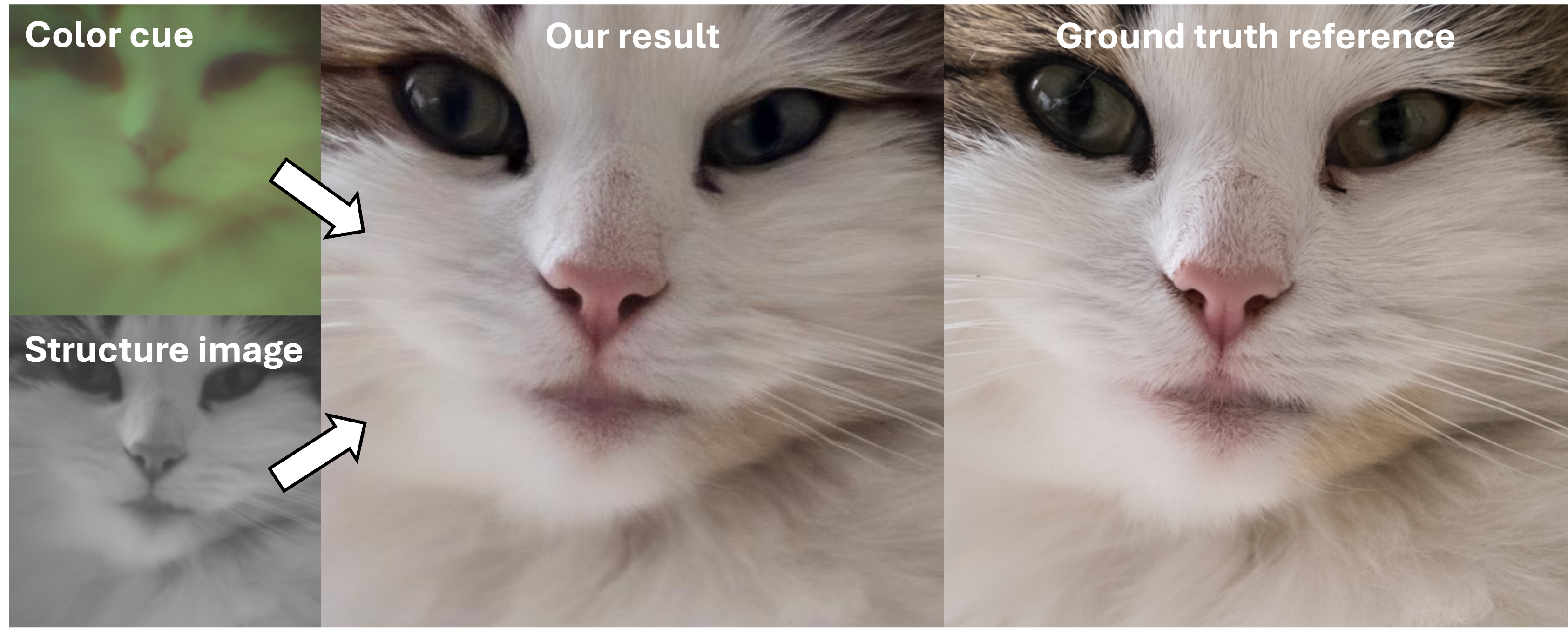}
\caption{Goal of this paper: We want to reconstruct a sharp color image from a pair of (a) structure image and (b) color cue. Shown in this figure are two \emph{real} images captured by a prototype metalens system we built, and (c) the reconstructed image. The focus of this paper is on the image reconstruction algorithm.}
\label{fig:fig1}
\end{figure}

The contribution of this paper is an algorithmic solution to overcome the spatially varying aberration problem that arises from metalenses. However, we emphasize that the method is not limited to metalenses. Any ordinary optical system with a similar chromatic aberration can use our proposed method.

As a preview of our solution, Figure~\ref{fig:fig1} presents the input image(s) and the reconstructed output. The input to our problem consists of a pair of images: (1) a chromatic-aberration–distorted color image, which is the native output of a metalens, and (2) a bandpass gray-scale structural image unaffected by chromatic aberration, which can be obtained by placing a bandpass filter in front of the same metalens. The exact optical implementation (i.e., how to construct the system with minimal increase in form factor) is omitted here, as it is beyond the scope of this algorithmic paper. Loosely speaking, a setup as simple as beam splitting is sufficient to achieve this goal.

The use of paired inputs is inspired by decades of research in hyperspectral pansharpening \cite{loncan2015hyperspectral, masi2016pansharpening, meng2023pandiff}, although here we focus specifically on the context of metalenses. Our intuition is that pixels suffering from severe chromatic aberration are beyond repair unless an additional source of information is provided. We address this by incorporating the monochrome structural image. By exploiting the color information already present in the distorted image and leveraging signal priors from diffusion models, we aim to recover the underlying image. To this end, we address two challenges in this paper:

\begin{enumerate}
\item \textbf{Dual-branch diffusion}: The color cue image and the structural image are intrinsically living in two different spaces with pixel misalignments. To ensure that we can extract meaningful signal from both, we introduce a \textit{spatial transformer} and a \textit{vision adaptor} to bring the two to the same latent space.
\item \textbf{Spatially varying blur conditioning}: The blur in our problem is spatially varying. Naive implementation of diffusion models cannot handle this type of blur. We introduce a new \textit{kernel prediction} method to estimate the blur at every pixel.
\end{enumerate}

%%%%%%%%%%%%%%%%%%%%%%%%%%%%%%%%%%%%%%%%%%%%%%%%%%%%%%%%%%%%%%%%%%%%%%%%%%%%%%%%%%%%%%%%%%%%%%%%%%%%%%%%%%%%%%%%%%%%%
\section{Background}
\subsection{Problem setting}
\begin{figure}[htb]
    \centering
    \includegraphics[width=0.8\linewidth]{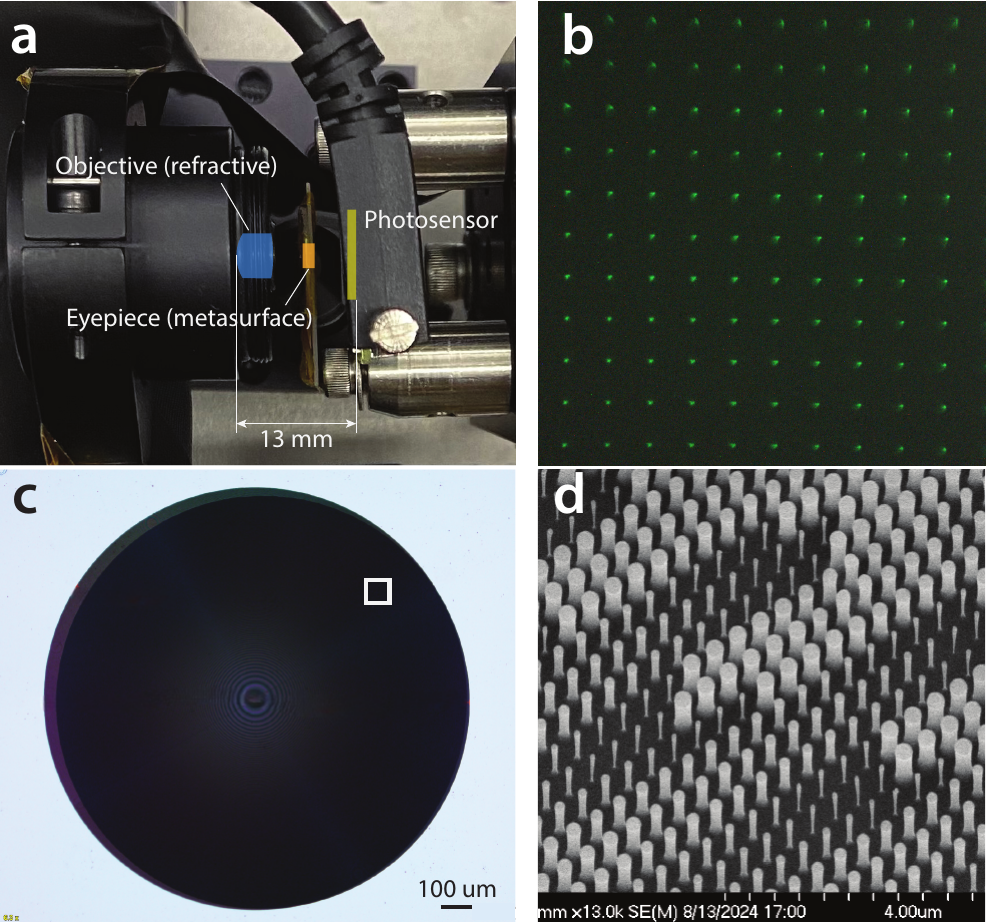}
    \caption{(a) MetaZoom optical assembly. It consists of the Thorlabs AC050-008-A-ML lens~(f=7.5 mm, Ø5 mm) as the objective, a custom metasurface as the eyepiece, and a Basler daA3840-45uc RGB as the photosensor. A 532~nm, 10~nm FWHM spectral filter can be inserted into the system to capture the structure image. (b) Measured PSFs of the structure image with a diagonal field of view of $5^{\circ}$.  (c) A sample metasurface under the optical microscope. (d) SEM image of the zoomed-in region of (c) at 13000x magnification.}
    \label{fig:optical_setup}
\end{figure}

The two images captured by the imaging system illustrated in \cref{fig:optical_setup} can be formulated by following two equations:
\begin{align*}
\vy_{\text{c}} &= \vh_c \overset{\text{s.v.}}{\circledast} \vx + \mathcal{N}(0,\mathbf{I}),\\
\vy_{\text{s}} &= \vh_s \overset{\text{s.v.}}{\circledast} (\mathbf{S}\vx) + \mathcal{N}(0,\mathbf{I}),
\end{align*}
where $\vy_c\in\mathbb{R}^{3N}$ is the color cue image, $\vy_s\in\mathbb{R}^{N}$ is the structure image. Here, $\vh_c$ and $\vh_s$ represent the spatially varying blur operators, $\mathbf{S}\in\mathbb{R}^{N\times3N}$ is the color-channel averaging matrix that converts the true color image $\vx\in\mathbb{R}^{3N}$ to a monochromatic image. The operation $\overset{\text{s.v.}}{\circledast}$ stands for spatially varying convolution, where the convolution kernel varies with position. $\mathcal{N}(0,\mathbf{I})$ represents the zero-mean unit-variance Gaussian noise.

Our goal is to reconstruct the high-quality color image $\vx$ from two degraded measurements captured by an optical system with a metalens. The first measurement is the color cue, $\vy_c$, which consists of severe spatially varying blur from optical aberrations. The second is a monochrome structure image, $\vy_s$, captured from the same metalens optical system with a bandpass filter that has sharper high-frequency details but lacks color information. 

\subsection{Spatially varying blur}
Spatially invarying blur removal has been studied for half a century. For blind deblurring, most methods are based on alternating minimization \cite{fergus2006removing, xu2010two}, with many new deep-learning methods from vision transformers to diffusion models \cite{dong_tci_2021_dwdn, kong2025deblurdiff}. A key observation that is often overlooked is the ill-posedness of the joint estimation problem, where some work pointed out that it is better to estimate the blur kernel before attempting to recover the image \cite{levin2006blind, sanghvi2024kernel}

For spatially varying blur, the known results are much sparser. Early work in applied mathematics constructs the full blur matrix and resorts to optimization tools. \cite{chan1998total, nagy1994fast, chan2011single} Newer approaches, such as \cite{yanny_optica_2022_mwn}, proposed a customized solution for microscopes, which is not generalizable to other systems. Our empirical findings show that, with appropriate training data, some deep learning deblurring models can also produce reasonable results, even when the blur is spatially varying. However, their performance is limited by the training data.

\subsection{Color fusion}
Our proposed method is inspired by multispectral fusion. However, the key difference is that in multispectral fusion, images generally have mild spatially varying blur. Under this context, popular methods such as pansharpening \cite{masi2016pansharpening, loncan2015hyperspectral}, hyperspectral signal recovery \cite{arad2016sparse}, and spectral decomposition \cite{schmidt2020guide} are somewhat easier to employ. When the spatially varying blur is present, deblurring while simultaneously recovering the color has never been attempted before.

\section{Method}
We propose a novel generative fusion framework designed to reconstruct a high-quality color image from two imperfect, degraded inputs. Our method corrects these optical aberrations by 1) aligning and fusing the sharp, high-frequency details from the structure image with the color information from the distorted color cue and 2) generating image information that was lost due to severe aberrations using a diffusion model.

\begin{figure}[htb]
    \centering
    \includegraphics[width=\linewidth]{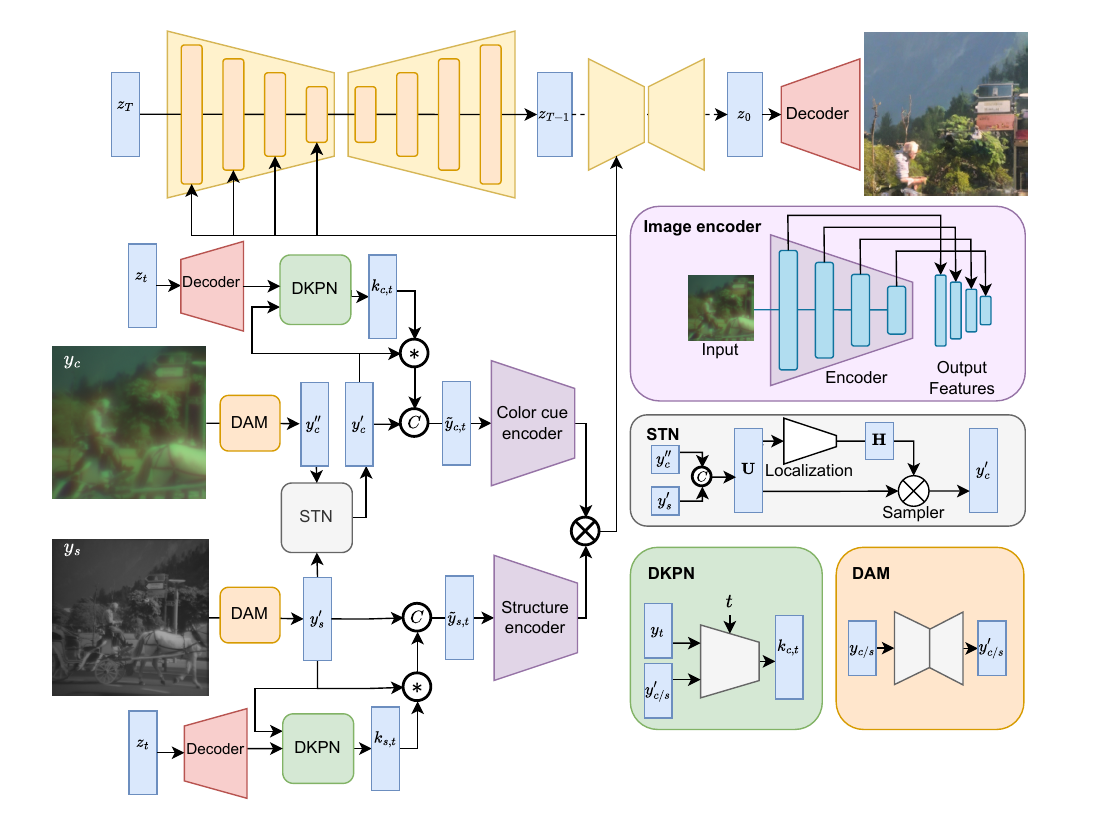}
    \caption{A schematic of the proposed framework for optical aberration mitigation and image fusion. The model is designed to reconstruct a high-quality color image from two imperfect, degraded inputs. It utilizes a dual-branch diffusion model built upon a pre-trained diffusion model \cite{podell_2023_sdxl}}
    \label{fig:postprocessing}
\end{figure}

Simply using the severely aberrated images to condition the diffusion network can produce poor results, as the model is forced to solve two problems simultaneously: deblurring the inputs and using them for image generation. First, let's discuss how the dual branch diffusion effectively fuses the two images by aligning the latents. 

\textbf{Domain adaptation.} Diffusion models are trained on tone-mapped images, while our measured inputs are in a different domain. Therefore, we first use a Domain Adaptation Module (DAM) to independently match both the structure and color images to the domain of the diffusion model, as illustrated in \cref{fig:postprocessing}. 

\textbf{Frame alignment.} In real-world applications, minor physical shifts in the optical system can cause the captured structural image and color information to become misaligned. This spatial misalignment can cause significant artifacts, most notably color bleeding at object edges. To resolve this, we first align the color cue to the structure image using a Spatial Transformer Network (STN)\cite{jaderberg2015spatial}. The STN predicts the geometric transformation $\mathbf{H}$ using a localization network as shown in \cref{fig:postprocessing}. This transformation is applied to the color cue in the image space, which ensures both images are spatially consistent before they are processed by the rest of the network.

\textbf{Vision adapter.} To provide the conditioning to the diffusion model, the image features of the color cue and the structure image are fused together using the concept of a \emph{vision adapter}, previously proposed by Mou et al. \cite{mou2024t2i}. The original adapter was developed for text-to-image generation with the goal of adding controllability to the generation process. 
In this work, we modify the original adapter so that it can perform the colorization task. 
The pre-trained stable diffusion XL~\cite{podell_2023_sdxl} takes the current estimate $\mathbf{z}_t$ and generates the next iterate $\mathbf{z}_{t-1}$. 
The structure of the diffusion block is a UNet with an encoder $\text{Encoder}_z(\cdot)$ and a decoder $\text{Decoder}_{z}(\cdot)$. The features extracted by the UNet's encoder are denoted as 
$\mathbf{f}_{z,t} = \text{Encoder}_z(\mathbf{z}_t),$ 
where $\mathbf{f}_{z,t}$ consists of multi-scale features extracted by the first part of the UNet.
Color encoder $\text{Encoder}_{\text{color}}(\cdot)$ is a trainable module aiming to extract features of the color image. These multi-scale features $\mathbf{f}_{c,t}$ have a similar dimensionality to the features extracted by the Unet. The structural encoder $\text{Encoder}_{\text{struc}}(\cdot)$ has a network architecture similar to $\text{Encoder}_\text{color}(\cdot)$.
\begin{figure*}[ht]
\begin{minipage}[b]{0.099\linewidth} \centerline{\includegraphics[width=\textwidth]{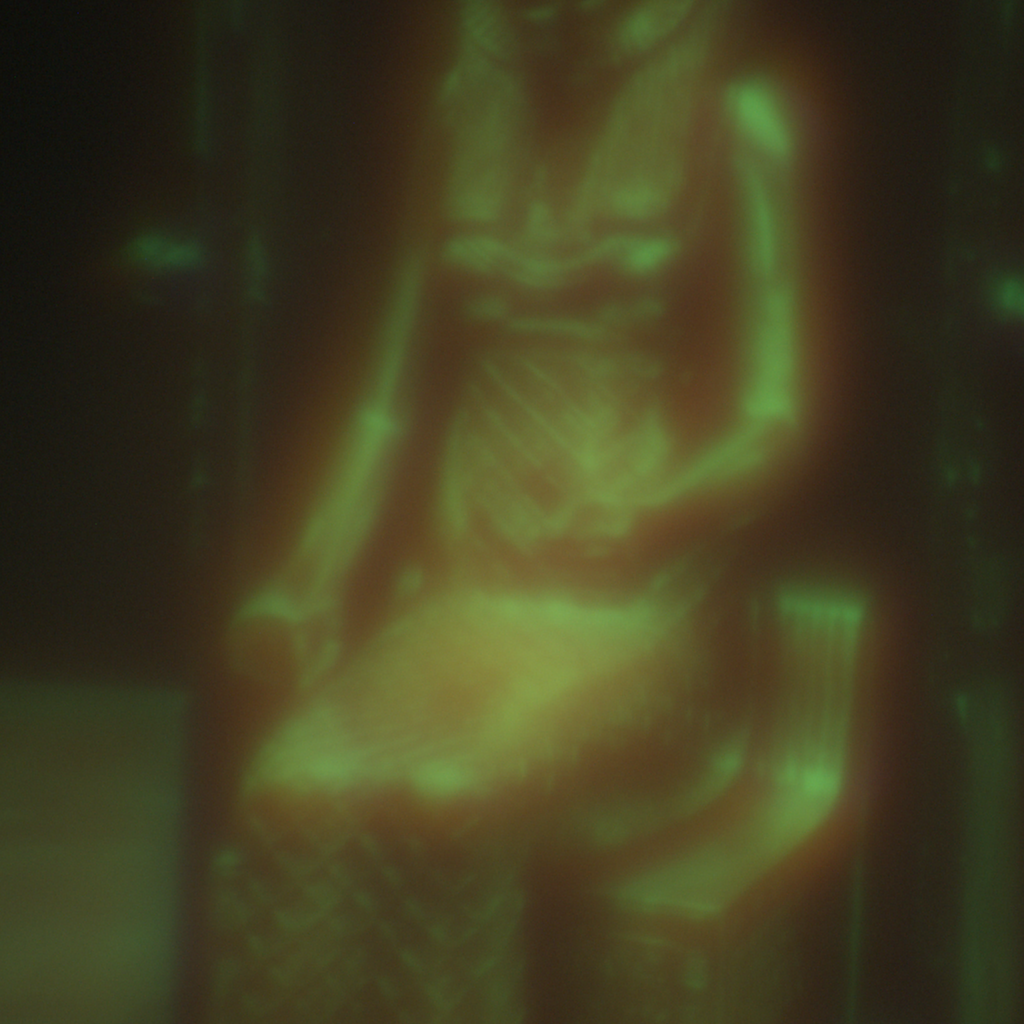}} \end{minipage} \hspace{-0.01\linewidth}
\begin{minipage}[b]{0.099\linewidth} \centerline{\includegraphics[width=\textwidth]{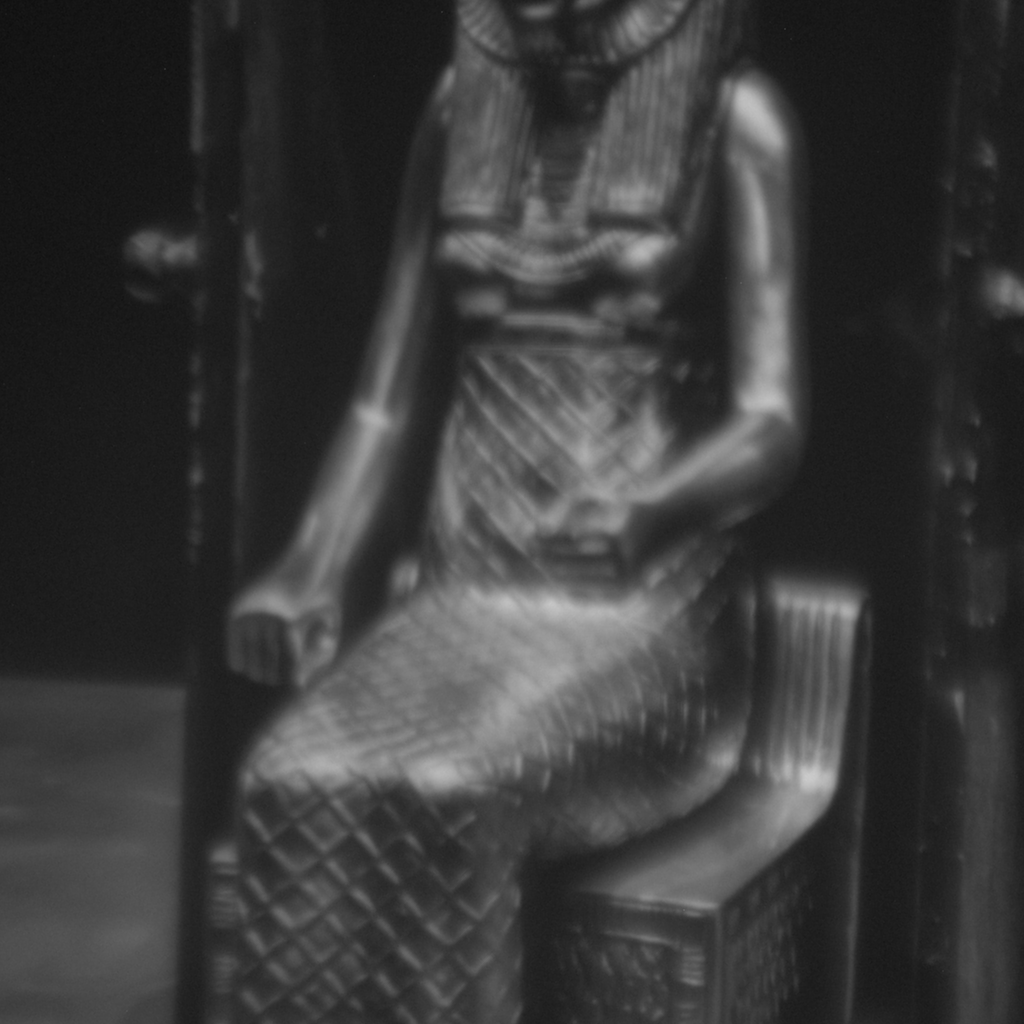}} \end{minipage} \hspace{-0.01\linewidth}
\begin{minipage}[b]{0.099\linewidth} \centerline{\includegraphics[width=\textwidth]{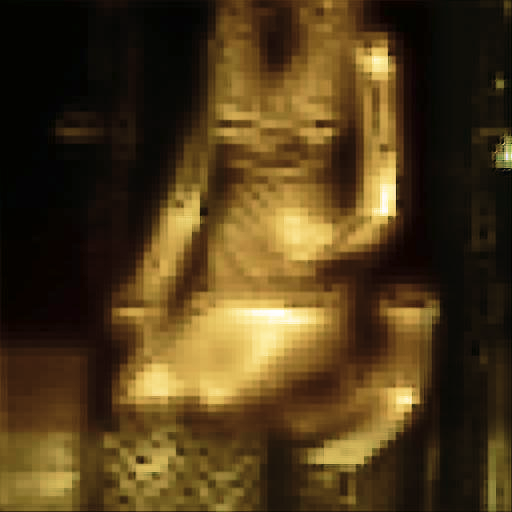}} \end{minipage} \hspace{-0.01\linewidth}
\begin{minipage}[b]{0.099\linewidth} \centerline{\includegraphics[width=\textwidth]{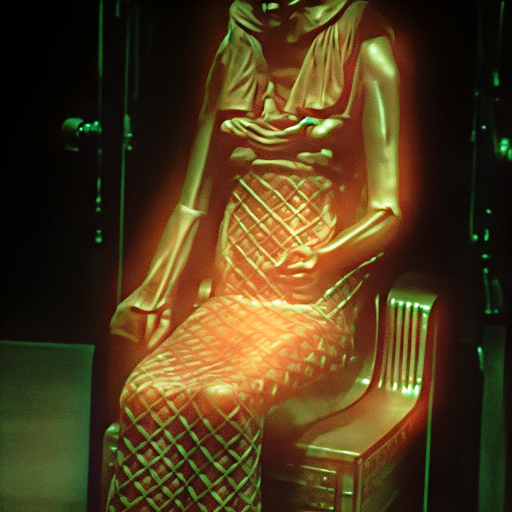}} \end{minipage} \hspace{-0.01\linewidth}
\begin{minipage}[b]{0.099\linewidth} \centerline{\includegraphics[width=\textwidth]{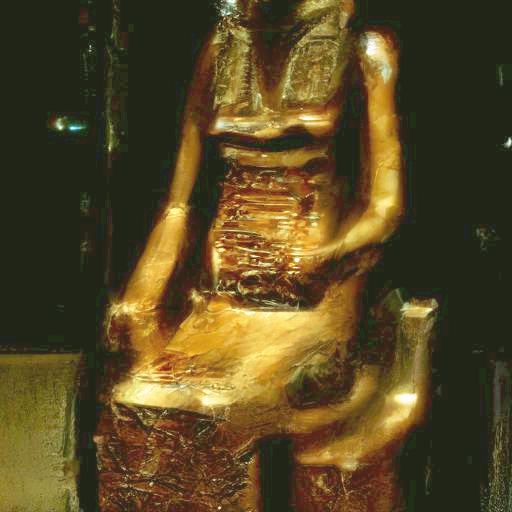}} \end{minipage} \hspace{-0.01\linewidth}
\begin{minipage}[b]{0.099\linewidth} \centerline{\includegraphics[width=\textwidth]{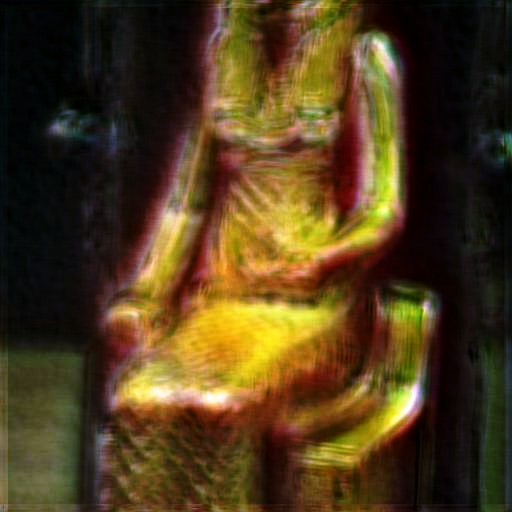}} \end{minipage} \hspace{-0.01\linewidth}
\begin{minipage}[b]{0.099\linewidth} \centerline{\includegraphics[width=\textwidth]{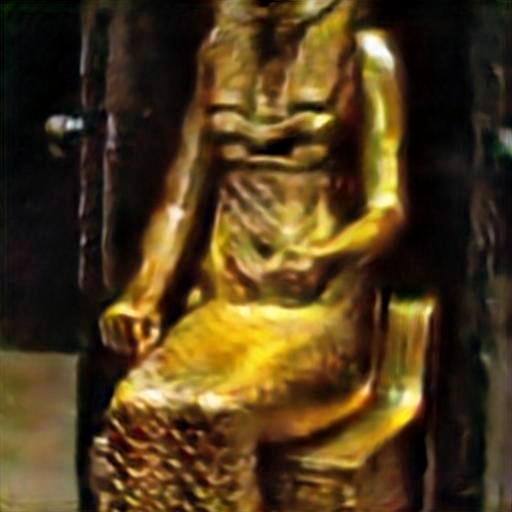}} \end{minipage} \hspace{-0.01\linewidth}
\begin{minipage}[b]{0.099\linewidth} \centerline{\includegraphics[width=\textwidth]{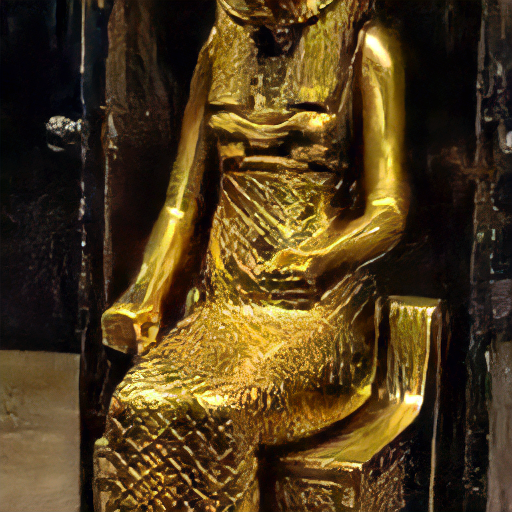}} \end{minipage} \hspace{-0.01\linewidth}
\begin{minipage}[b]{0.099\linewidth} \centerline{\includegraphics[width=\textwidth]{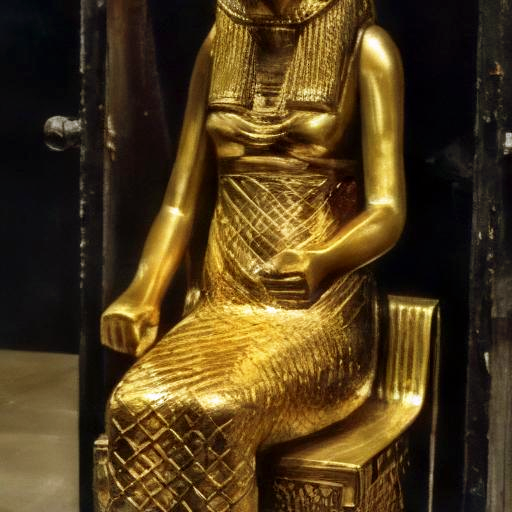}} \end{minipage} \hspace{-0.01\linewidth}
\begin{minipage}[b]{0.099\linewidth} \centerline{\includegraphics[width=\textwidth]{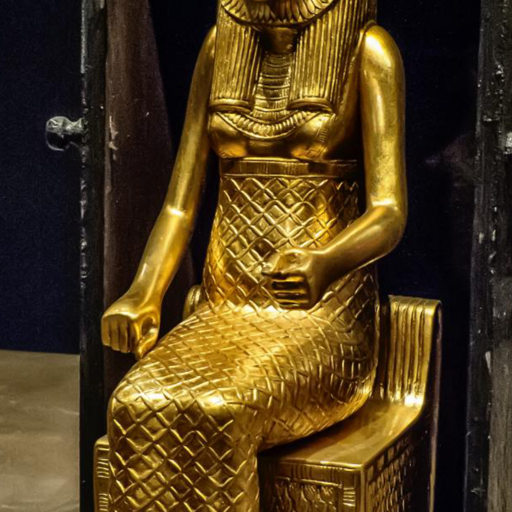}} \end{minipage} \hspace{-0.01\linewidth}

\begin{minipage}[b]{0.099\linewidth} \centerline{\includegraphics[width=\textwidth]{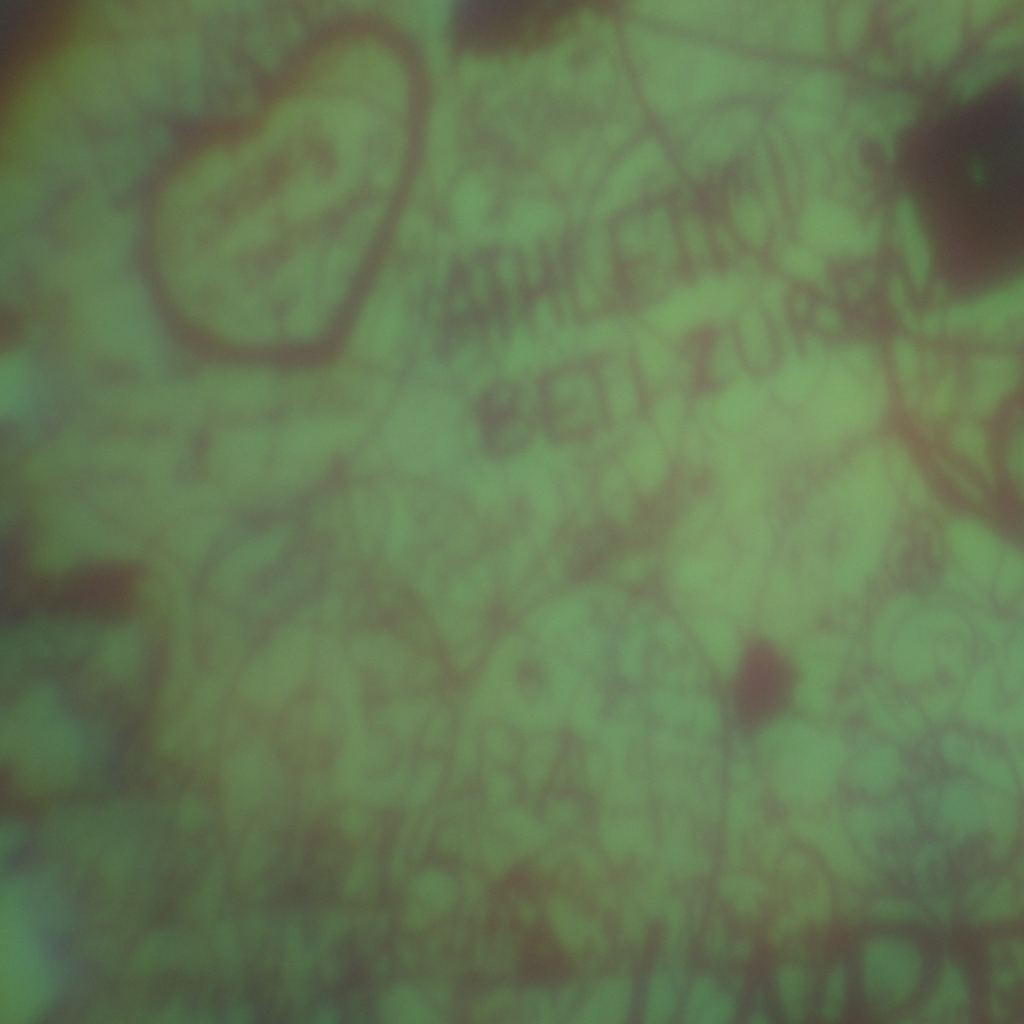}} \end{minipage} \hspace{-0.01\linewidth}
\begin{minipage}[b]{0.099\linewidth} \centerline{\includegraphics[width=\textwidth]{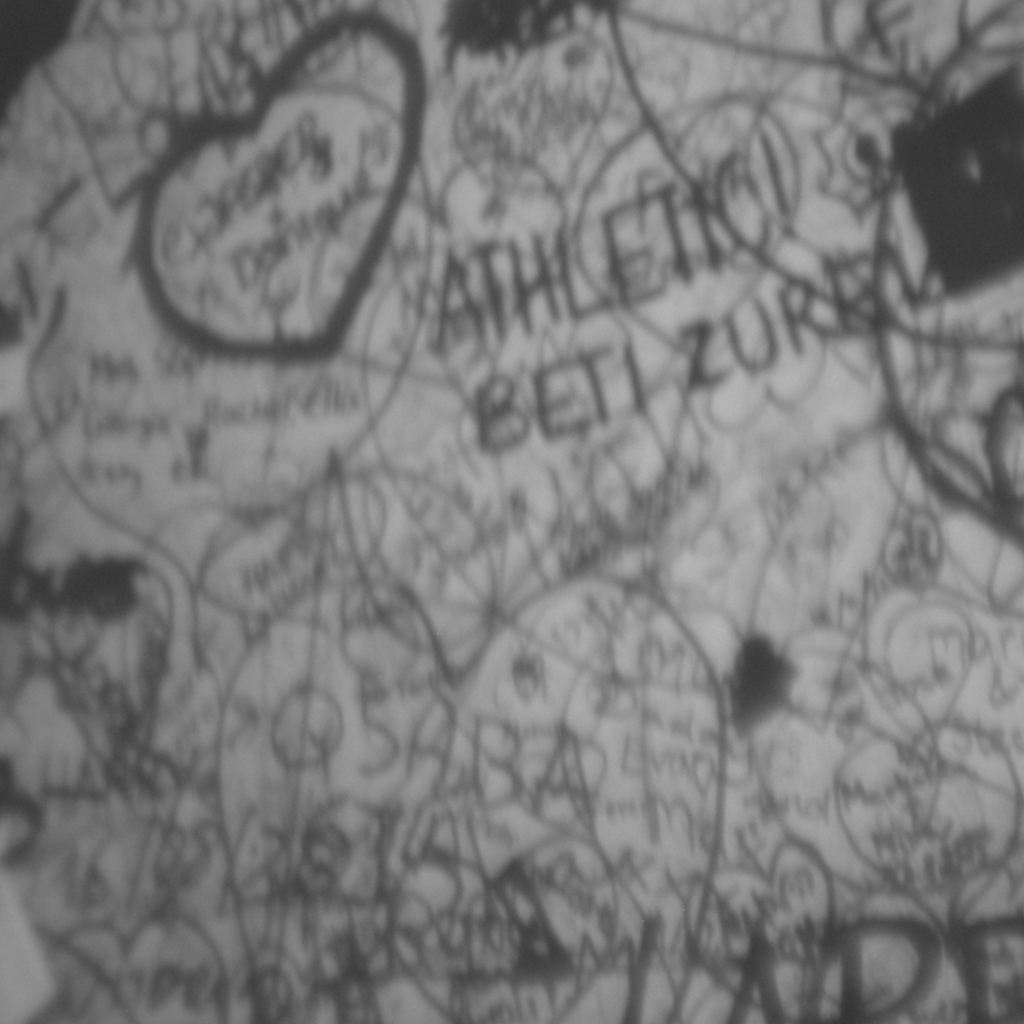}} \end{minipage} \hspace{-0.01\linewidth}
\begin{minipage}[b]{0.099\linewidth} \centerline{\includegraphics[width=\textwidth]{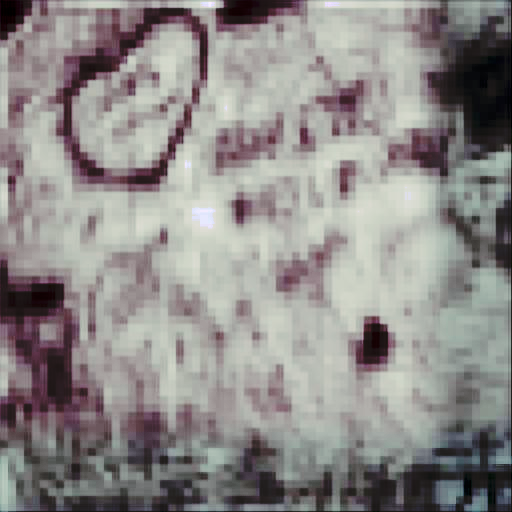}} \end{minipage} \hspace{-0.01\linewidth}
\begin{minipage}[b]{0.099\linewidth} \centerline{\includegraphics[width=\textwidth]{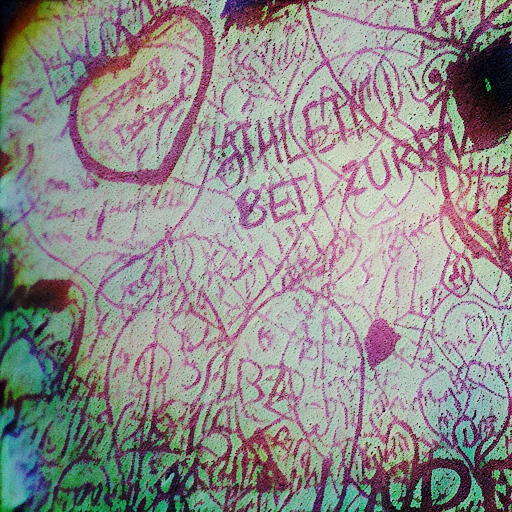}} \end{minipage} \hspace{-0.01\linewidth}
\begin{minipage}[b]{0.099\linewidth} \centerline{\includegraphics[width=\textwidth]{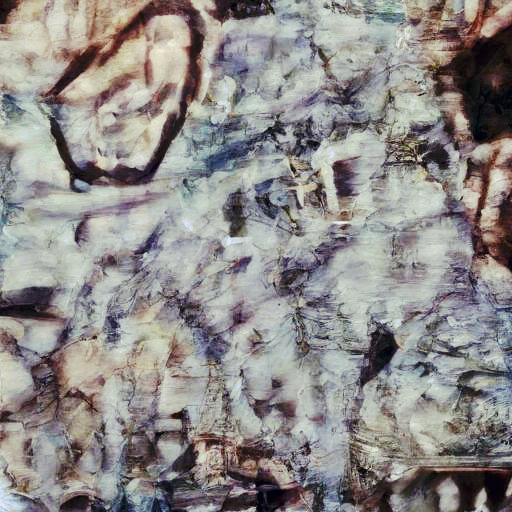}} \end{minipage} \hspace{-0.01\linewidth}
\begin{minipage}[b]{0.099\linewidth} \centerline{\includegraphics[width=\textwidth]{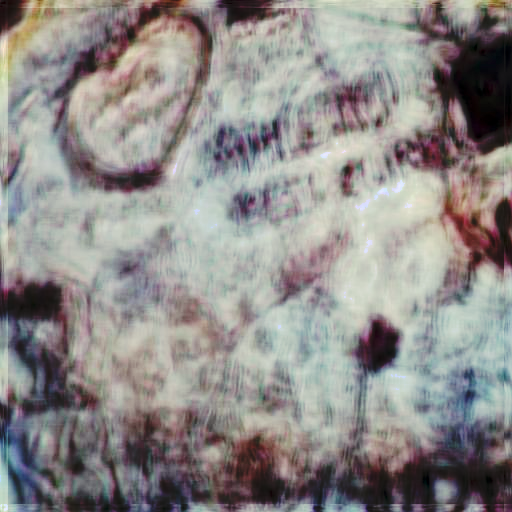}} \end{minipage} \hspace{-0.01\linewidth}
\begin{minipage}[b]{0.099\linewidth} \centerline{\includegraphics[width=\textwidth]{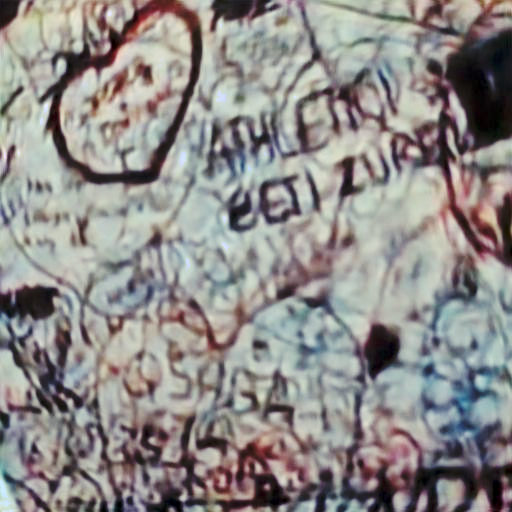}} \end{minipage} \hspace{-0.01\linewidth}
\begin{minipage}[b]{0.099\linewidth} \centerline{\includegraphics[width=\textwidth]{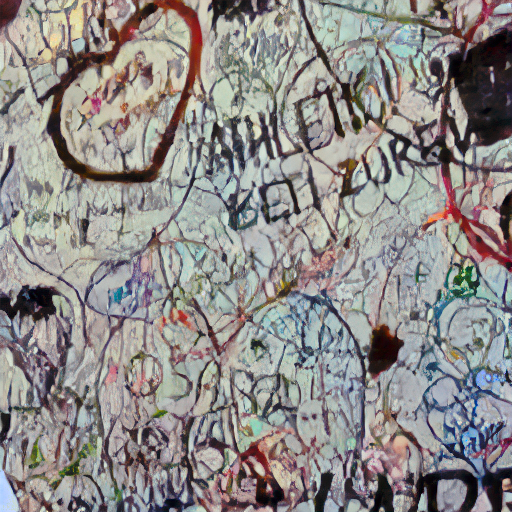}} \end{minipage} \hspace{-0.01\linewidth}
\begin{minipage}[b]{0.099\linewidth} \centerline{\includegraphics[width=\textwidth]{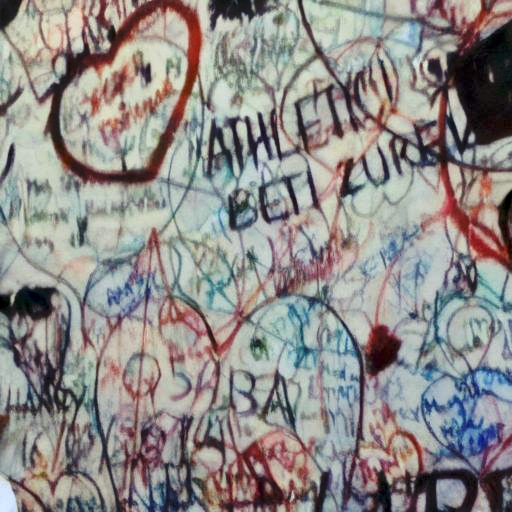}} \end{minipage} \hspace{-0.01\linewidth}
\begin{minipage}[b]{0.099\linewidth} \centerline{\includegraphics[width=\textwidth]{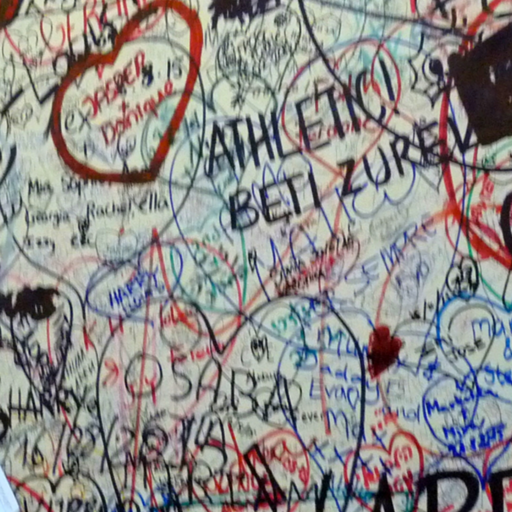}} \end{minipage} \hspace{-0.01\linewidth}

\begin{minipage}[b]{0.099\linewidth} \centerline{\includegraphics[width=\textwidth]{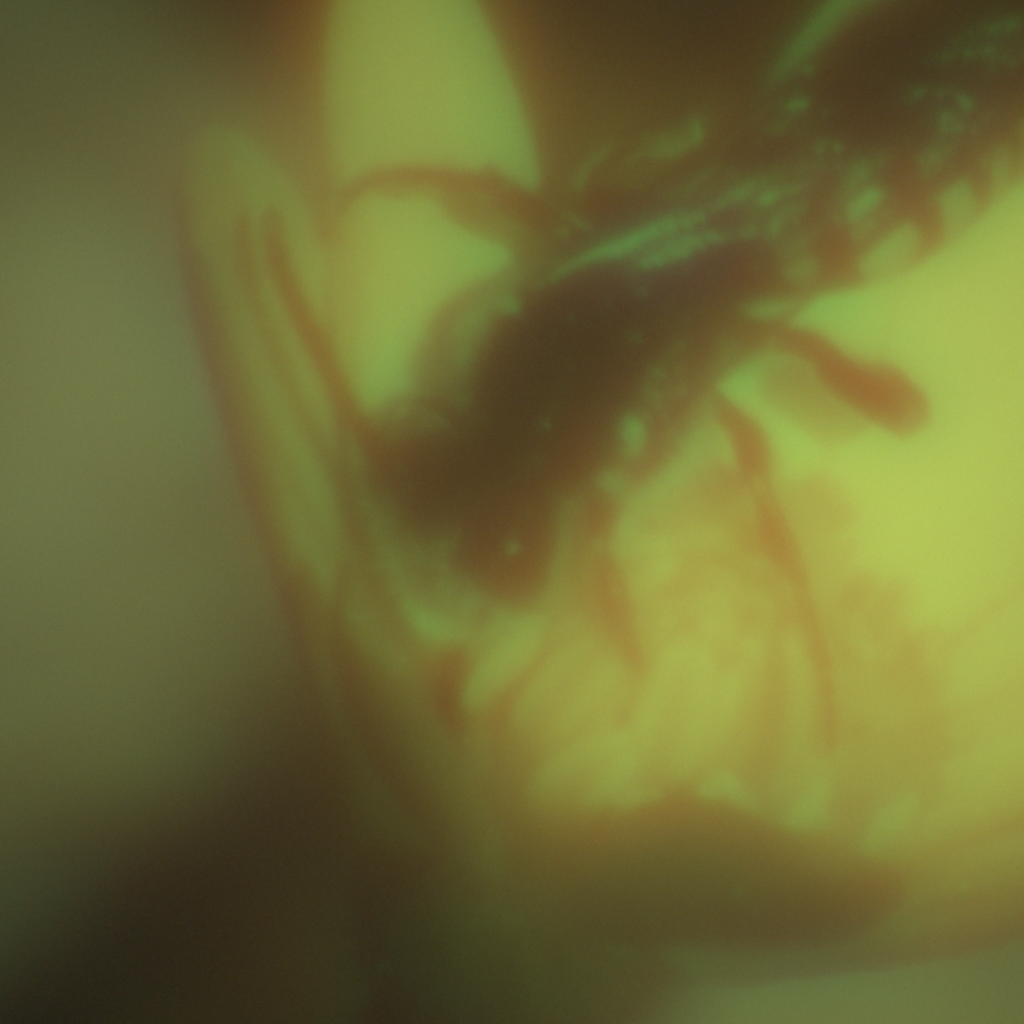}} \end{minipage} \hspace{-0.01\linewidth}
\begin{minipage}[b]{0.099\linewidth} \centerline{\includegraphics[width=\textwidth]{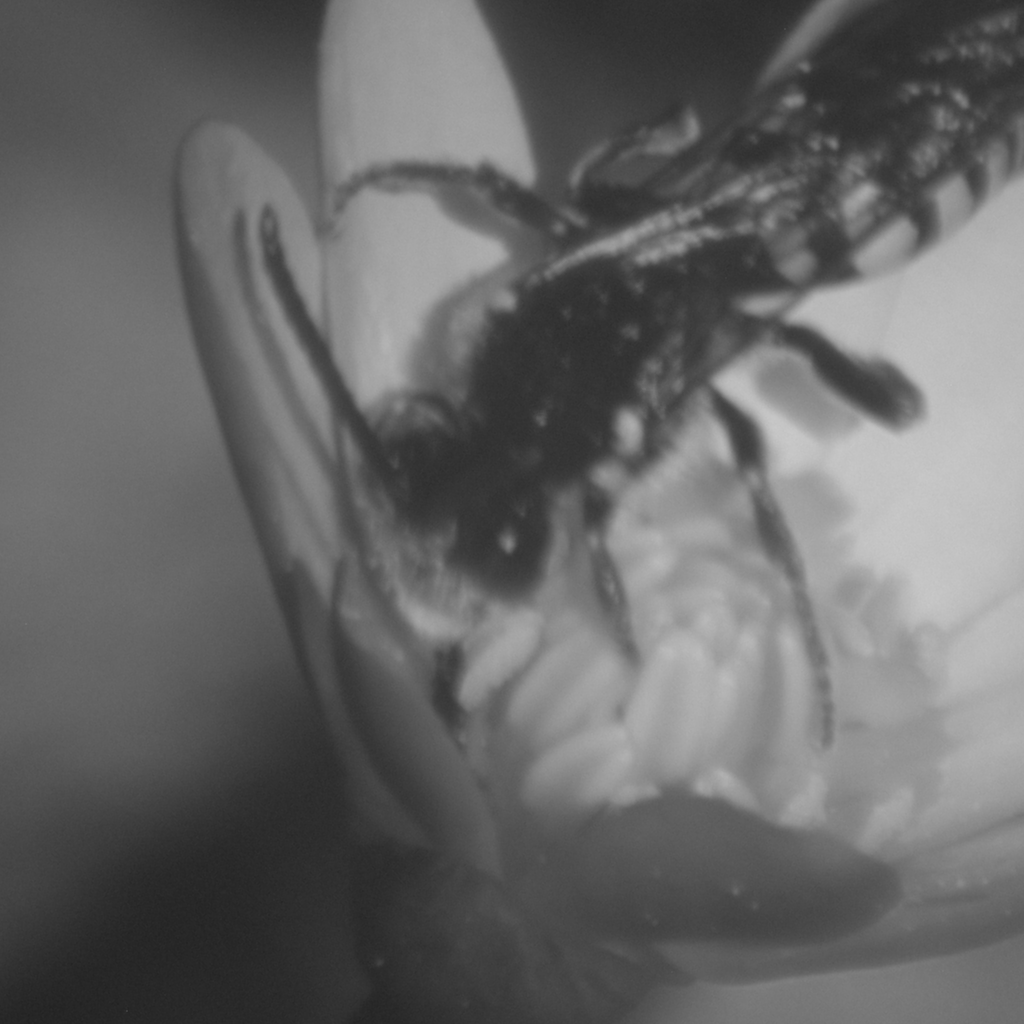}} \end{minipage} \hspace{-0.01\linewidth}
\begin{minipage}[b]{0.099\linewidth} \centerline{\includegraphics[width=\textwidth]{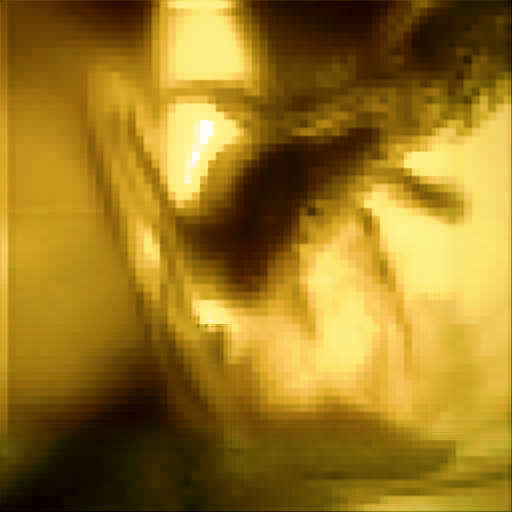}} \end{minipage} \hspace{-0.01\linewidth}
\begin{minipage}[b]{0.099\linewidth} \centerline{\includegraphics[width=\textwidth]{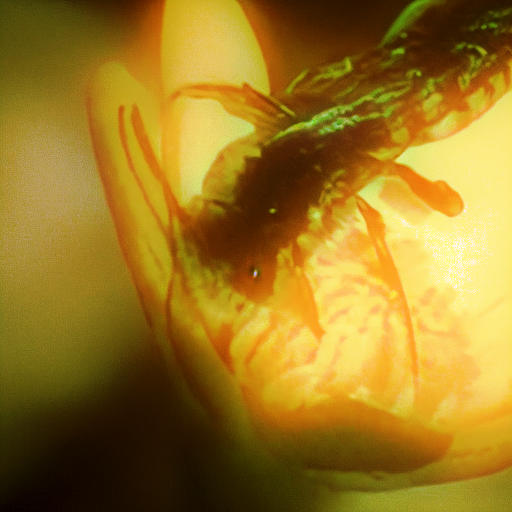}} \end{minipage} \hspace{-0.01\linewidth}
\begin{minipage}[b]{0.099\linewidth} \centerline{\includegraphics[width=\textwidth]{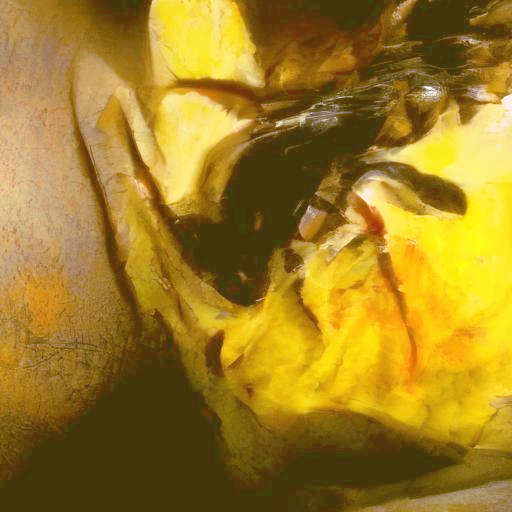}} \end{minipage} \hspace{-0.01\linewidth}
\begin{minipage}[b]{0.099\linewidth} \centerline{\includegraphics[width=\textwidth]{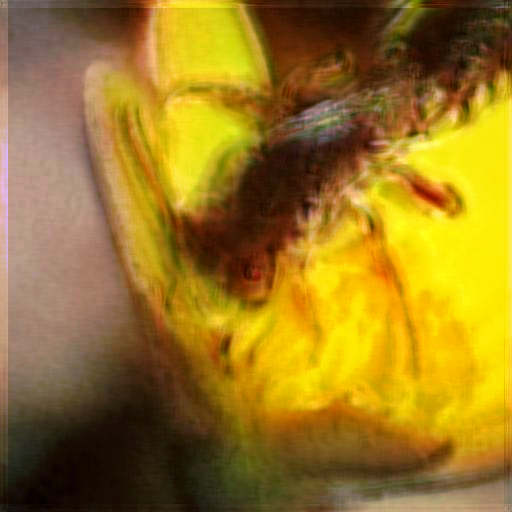}} \end{minipage} \hspace{-0.01\linewidth}
\begin{minipage}[b]{0.099\linewidth} \centerline{\includegraphics[width=\textwidth]{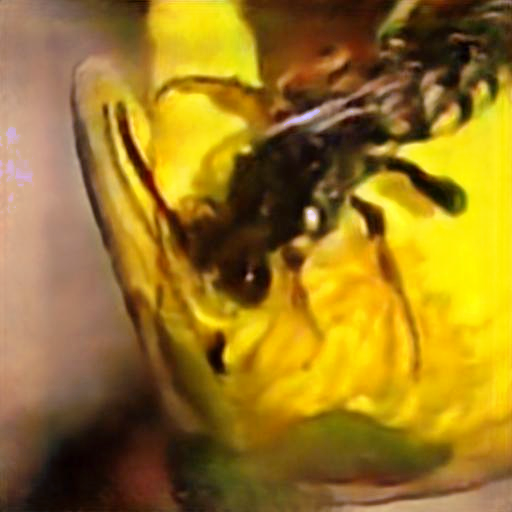}} \end{minipage} \hspace{-0.01\linewidth}
\begin{minipage}[b]{0.099\linewidth} \centerline{\includegraphics[width=\textwidth]{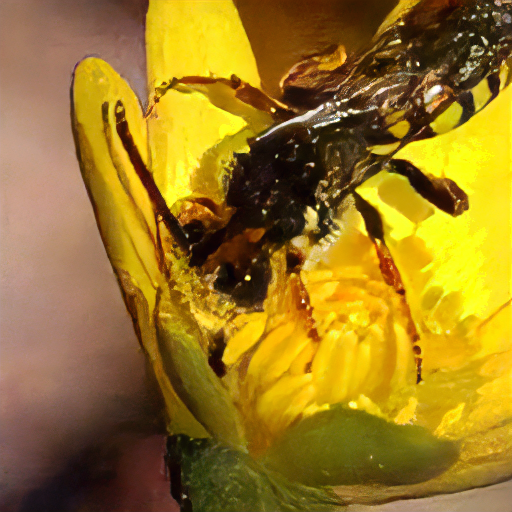}} \end{minipage} \hspace{-0.01\linewidth}
\begin{minipage}[b]{0.099\linewidth} \centerline{\includegraphics[width=\textwidth]{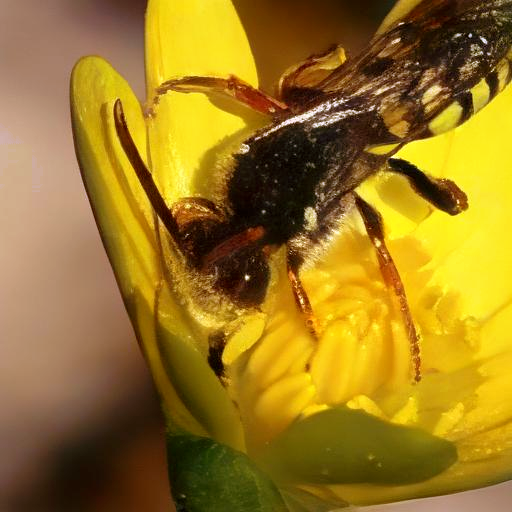}} \end{minipage} \hspace{-0.01\linewidth}
\begin{minipage}[b]{0.099\linewidth} \centerline{\includegraphics[width=\textwidth]{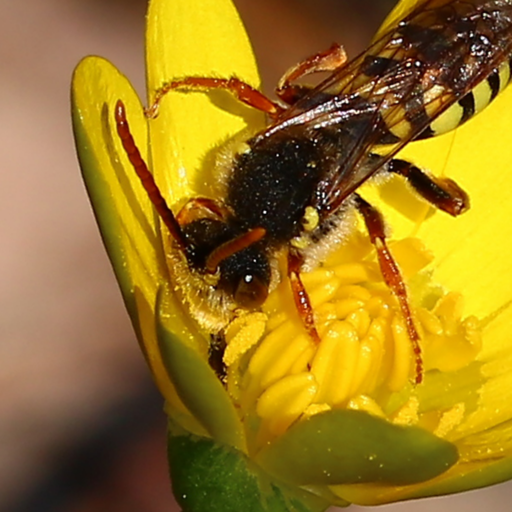}} \end{minipage} \hspace{-0.01\linewidth}

\begin{minipage}[b]{0.099\linewidth} \centerline{$\vy_c$}\medskip \end{minipage} \hspace{-0.01\linewidth}
\begin{minipage}[b]{0.099\linewidth} \centerline{$\vy_s$}\medskip \end{minipage} \hspace{-0.01\linewidth}
\begin{minipage}[b]{0.099\linewidth} \centerline{Unet}\medskip \end{minipage} \hspace{-0.01\linewidth}
\begin{minipage}[b]{0.099\linewidth} \centerline{DiffBIR}\medskip \end{minipage} \hspace{-0.01\linewidth}
\begin{minipage}[b]{0.099\linewidth} \centerline{DeblurDiff}\medskip \end{minipage} \hspace{-0.01\linewidth}
\begin{minipage}[b]{0.099\linewidth} \centerline{PNN}\medskip \end{minipage} \hspace{-0.01\linewidth}
\begin{minipage}[b]{0.099\linewidth} \centerline{SRPPNN}\medskip \end{minipage} \hspace{-0.01\linewidth}
\begin{minipage}[b]{0.099\linewidth}  \centerline{ResShift}\medskip \end{minipage} \hspace{-0.01\linewidth}
\begin{minipage}[b]{0.099\linewidth}  \centerline{Ours}\medskip \end{minipage} \hspace{-0.01\linewidth}
\begin{minipage}[b]{0.099\linewidth} \centerline{GT}\medskip \end{minipage} \hspace{-0.01\linewidth}

\caption{Qualitative comparison with prior work using the real data captured from the metalens optical system. }
\label{fig:comparison_real}
\end{figure*}

The color and structural features are fused according to a simple addition rule:
% \begin{equation*}
    $\widehat{\mathbf{f}}_t^{i} = \mathbf{f}_{z,t}^{i} + \mathbf{f}_{c,t}^{i} \cdot \mathbf{f}_{s,t}^{i}$.
% \end{equation*}
This multiplicative fusion acts as a gating mechanism. The structure features ($\mathbf{f}_{s,t}$), which encode high-frequency details, effectively control where the color features ($\mathbf{f}_{c,t}$) are applied. This ensures that strong color information is only fused in areas with high structural confidence, preventing artifacts like color bleeding across sharp boundaries.
This fused feature is then sent to the diffusion UNet $\text{Encoder}_z$ to alter the input features of the diffusion. The decoded signal is
$\mathbf{z}_{t-1} = \text{Decode}_z(\widehat{\mathbf{f}}_t^1,\ldots,\widehat{\mathbf{f}}_t^4)$.

\textbf{Pre-deblurring.} We use a Deblurring Kernel Prediction Network (DKPN) to estimate the spatially varying blur and produce an initial, coarse deblurred version of the image. This pre-processing step allows the main diffusion model to focus on its primary task: refining details and accurately fusing the two image sources. 

These predicted kernels ($\mathbf{k}_{c,t}$) are convolved with the domain-mapped blurred image to generate an initial estimate of the deblurred image. 

For each input, the deblurred image is concatenated with the original image. This provides the encoders with both a strong initial estimate of the deblurred structure and the unaltered low-frequency color information from the original input, ensuring no crucial details are lost.

\textbf{Training losses.}
Our framework is trained end-to-end with a composite loss function that addresses the two key stages of the process: initial deblurring and diffusion-based refinement. 
To ensure the DKPN modules produce a useful initial reconstruction, we calculate a per-pixel loss between the deblurred color cue $\tilde{\mathbf{y}}_{c,t}$ and the ground truth $\mathbf{x}$ using MSE loss. Similarly, MSE loss is calculated between $\tilde{\mathbf{y}}_{s,t}$ and $\mathbf{x}$ and summed together to get the loss $\mathcal{L}_\text{DKPN}$ to update $\phi$ parameters of the DKPN module and $\mu$ parameters of the STN module.
\begin{equation*}
    \mathcal{L}_\text{DKPN} = \mathcal{L}_\text{MSE}(\tilde{\mathbf{y}}_{c,t}, \mathbf{x}) + \mathcal{L}_\text{MSE}(\tilde{\mathbf{y}}_{s,t}, \mathbf{x})
\end{equation*}

For the main network, we use the standard diffusion model loss. At each timestep, the network is trained to predict the noise that was added to the image. 
\begin{equation*}
    \mathcal{L}_\text{diff} = \mathbb{E}_{t \sim U(1,T), \vx_0 \sim p_\text{data}, \epsilon \sim \mathcal{N}(0,\mathbf{I})}\left[ \|\epsilon - \epsilon_\theta(\mathbf{z}_t, \mathbf{f}_{c,t}, \mathbf{f}_{s,t}, t) \|^2 \right]
\end{equation*}
where $\vx_0$ is the ground truth image sampled from the data distribution $p_\text{data}$, $\epsilon$ is the Gaussian noise, $\epsilon_\theta$ is the diffusion network with $\theta$ parameters. These $\theta$ parameters are updated using aforementioned $\mathcal{L}_\text{diff}$ loss.

\section{Experiments}
We compare our method with leading deblurring techniques and pansharpening algorithms to demonstrate its unique strengths. 
Our results highlight a key trade-off in modern image restorations: fidelity versus perceptual quality \cite{blau2018perception}. 
As \cref{tab:comparison_real}, our diffusion-based approach achieves a state-of-the-art FID score. This indicates that it excels at generating perceptually convincing images with plausible high-frequency details. Our method is designed to produce results that are not just numerically accurate but also visually compelling.

We retrained existing blind deblurring methods \cite{yue2023resshift, kong2025deblurdiff} using our blur dataset with optical aberrations to make it a fair comparison. But these methods show poor performance compared to our method. Since the core idea behind our problem also relates to colorization, we compare our method with existing pansharpening methods such as PNN \cite{masi2016pansharpening} and SRPPNN \cite{cai2020srppnn}. Since these methods are finetuned to a particular dataset, we retrained these methods on our data. However, these methods also generate suboptimal results due to a lack of deblurring capability in their architectures. Our method shows superior feature fusion by colorizing the image while improving the structural features.

\begin{table}[ht]
    \centering
    \begin{tabular}{lccc}
    \toprule
    \textbf{Method} & \textbf{PSNR $\uparrow$} & \textbf{LPIPS $\downarrow$} & \textbf{FID $\downarrow$} \\ \hline
    Unet & 16.3306 & 0.6138 & 21.7120 \\ 
    % MPRNet &  &  &  \\ 
    ResShift & 21.4616 &	0.4158	& 0.7780 \\ 
    DiffBIR & 14.0906 & 0.5928 & 12.9938 \\ 
    DeblurDiff & 17.8541 & 0.5473 & 3.0188 \\ 
    PNN & 17.4618 & 0.5588 & 15.2630 \\ 
    SRPNN & 20.7870 & 0.4880 & 13.0580 \\ 
    Ours  & \textbf{22.5997} & \textbf{0.3184} & \textbf{1.8516} \\ 
    \bottomrule
    \end{tabular}
    \caption{Performance comparison of different methods in terms of PSNR, LPIPS, and FID.}
    \label{tab:comparison_real}
\end{table}

\subsection{Ablation study}
To verify the efficacy of using the proposed modules in our method, we retrained our method without the adaptation modules. A qualitative comparison is given in \cref{tab:comparison_ablation}, where it shows that the proposed modules increase the performance overall. 

\begin{table}[ht]
    \centering
    \begin{tabular}{lccc}
        \toprule
        \textbf{Method} & \textbf{PSNR ↑} & \textbf{LPIPS ↓} & \textbf{FID ↓} \\ \hline
        % Ours - without DKPN &  &  &  \\ 
        Ours - without STN & 20.9569 & 0.4042 & 1.0579 \\ 
        Ours - without DAM & 18.5551 & 0.4016 & 3.3786 \\ 
        Ours & \textit{22.5997} & \textit{0.3184} & \textit{1.8516} \\ 
        % Ours (10\% structure) &  &  &  \\ 
        % Ours (10\% color cue) &  &  &  \\ 
        \bottomrule
    \end{tabular}
    \caption{Ablation study of the proposed method.}
    \label{tab:comparison_ablation}
\end{table}
\vspace{-10pt}
\section{Conclusion}
We introduced a generative image fusion framework that effectively corrects severe, spatially varying aberrations in metalens imaging systems. By intelligently combining a sharp monochrome image with a distorted color cue, our diffusion-based method produces high-quality, perceptually realistic color images where traditional deblurring and fusion techniques fail.

\vfill\pagebreak

\footnotesize
\bibliographystyle{IEEEbib}
\bibliography{strings,refs}

\end{document}